\documentclass[conference,11pt]{IEEEtran}

\usepackage[utf8]{inputenc}
\usepackage[T1]{fontenc}
\usepackage{graphicx}
\usepackage{amsmath,amssymb,amsfonts}
\usepackage{booktabs}
\usepackage{hyperref}
\usepackage{subcaption}
\usepackage{xcolor}
\usepackage{url}
\usepackage{caption}
\usepackage{cite}
\usepackage{etoolbox}

\hypersetup{
  colorlinks=true,
  linkcolor=black,
  citecolor=black,
  urlcolor=black,
}

\begin{document}

\title{Self-Supervised Training For Low-Dose CT Reconstruction}

\author{
\IEEEauthorblockN{Mehmet Ozan Unal\textsuperscript{1}, Metin Ertas\textsuperscript{2}, Isa Yildirim\textsuperscript{1}}
\\
\IEEEauthorblockA{\textsuperscript{1}Electronics and Communication Engineering Department, Istanbul Technical University, Istanbul, Turkey\\
\textsuperscript{2}Electrical and Electronics Engineering Department, Istanbul University, Istanbul, Turkey\\
Email: unalmehmet@itu.edu.tr, ertas@istanbul.edu.tr, iyildirim@itu.edu.tr}
}

\maketitle

\begin{abstract}

Ionizing radiation has been the biggest concern in CT imaging. To reduce the dose level without compromising the image quality, low-dose CT reconstruction has been offered with the availability of compressed sensing based reconstruction methods. Recently, data-driven methods got attention with the rise of deep learning, the availability of high computational power, and big datasets. Deep learning based methods have also been used in low-dose CT reconstruction problem in different manners. Usually, the success of these methods depends on labeled data. However, recent studies showed that training can be achieved successfully with noisy datasets. In this study, we defined a training scheme to use low-dose sinograms as their own training targets. We applied the self-supervision principle in the projection domain where the noise is element-wise independent which is a requirement for self-supervised training methods. Using the self-supervised training, the filtering part of the FBP method and the parameters of a denoiser neural network are optimized. We demonstrate that our method outperforms both conventional and compressed sensing based iterative reconstruction methods qualitatively and quantitatively in the reconstruction of analytic CT phantoms and real-world CT images in low-dose CT reconstruction task.

\end{abstract}


\section{Introduction}

X-ray Computed Tomography (CT) uses ionizing radiation to noninvasively monitorize the human body. Ionizing radiation can be harmful to human body, therefore reducing the radiation dose without sacrificing its imaging quality is crucial. Recently, deep learning (DL) has become an alternative to solve low-dose CT problem. 
Although DL based methods have promising results, its success usually depends on sufficient amount of labeled datasets. Besides, unlike the natural image domain, evaluating the quality of the datasets requires domain experts such as radiologists. Therefore, solving this problem with classical DL methods is both cost and time intensive. However, recently it was shown that successful training could also be possible with noisy labels for image denoising problems \cite{noise2noise}. It was also shown that noisy images could be used as their own training targets \cite{noise2self,noise2void,self2selfdropout}. However, these methods were designed to be used in image denoising problems. In this manner, we extended the self-supervised approach of these methods for low-dose CT reconstruction problem. Therefore, we designed an algorithm which applies the self-supervision in the sinogram domain with the help of differentiable backward and forward operators of CT reconstruction. Our method aims to optimize the filtering part of the filtered back projection (FBP) method and learn the parameters of the denoiser neural network via self-supervised training.  Our method is realized in three different manners: \textit{i}) single sinogram self-supervised, \textit{ii}) learned single shot, \textit{iii}) learned self-supervised.

\section{Related Work}

A classical supervised training procedure for denoising problems can be defined as:

\begin{equation}
    \label{equ:supervised}
    \theta^{*}=\arg \min _{\theta} \mathbb{E}_{ x}[\|f_{\theta}(x)-y\|_{2}^{2}]
\end{equation}

where $\mathbb{E}_{x}$ is the expected value over $x$, ($x$, $y$) are paired samples: $x$ is the noisy measurements and $y$ is the noise-free measurements of the same information, $f_{\theta}$ is a deep neural network parameterized by $\theta$. During the optimization process, a function that maps noisy images to noise-free images is learned. Recently proposed Noise2Noise method claimed that supervised training is possible with only noisy measurements given that the noise is independent and additive. In addition, it was also shown that training with noisy image pairs should give similar results to training with noisy-clean image pairs \cite{noise2noise}.


 
Noise2Noise method enabled training through noisy targets but still, two noisy measurements of the same information are required. Noise2Self is a method that enables training with just noisy measurements via self-supervision \cite{noise2self}. The study proposed that noisy image itself can be used as its own target. Since the neural network can minimize the loss function just by converging to an identity function, to solve this issue, Noise2Self method suggested a masking mechanism that perturbs the image according to certain rules and prevents the neural network to converge identity function which is called J invariant. The proposed cost function by Noise2Self is formulated as follows:


\begin{equation}
    \label{equ:n2s_jinv}
    \theta^{*}=\arg \min _{\theta} \mathbb{E}_{ x}[\|f_{\theta}(x_{Jc})-x_{J}\|_{2}^{2}]
\end{equation}

where $J$ is a subset of pixels, $x_{J}$ is the $J^{th}$ subset of $x$, $x_{Jc}$ is the modified form of $x$ in such a way that $J^{th}$ pixels of $x$ are modified using the pixel values of $x$ except $J^{th}$. In other words, $J^{th}$ subset of $x$ is perturbed using any pixel other than $J^{th}$ subset. It is used as the input of the denoiser ($f_{\theta}$) and the reconstruction loss is only calculated for the $J^{th}$ subset of the pixels. Noise2Self study showed that training with self-supervised loss should result in similar to training with noise-free targets if the following conditions are met: \textit{i)} $x$ and $y$ should be conditionally independent. \textit{ii)} noise should be element-wise independent.

The opportunity of training on noisy labels could be quite valuable for medical inverse problems such as low-dose CT reconstruction. However, Noise2Self method may not be applied in a straight-forward way on low-dose CT reconstruction problem, since the noise and artifacts, which are formed during the reconstruction process, are not element-wise independent. Therefore, we developed a method which applies self-supervision on the projection domain where the noise can be modeled as independent. There is also a method which applies self-supervision on a single image and tries to learn an image domain denoiser \cite{noise2inverse}. However, in our approach, we used self-supervision to train both a frequency-domain filter that optimizes the filtering part of FBP and a neural network that denoises the reconstruction.

\section{Method}

\begin{figure}[th]
    \begin{center}
        \includegraphics[width=0.45\textwidth]{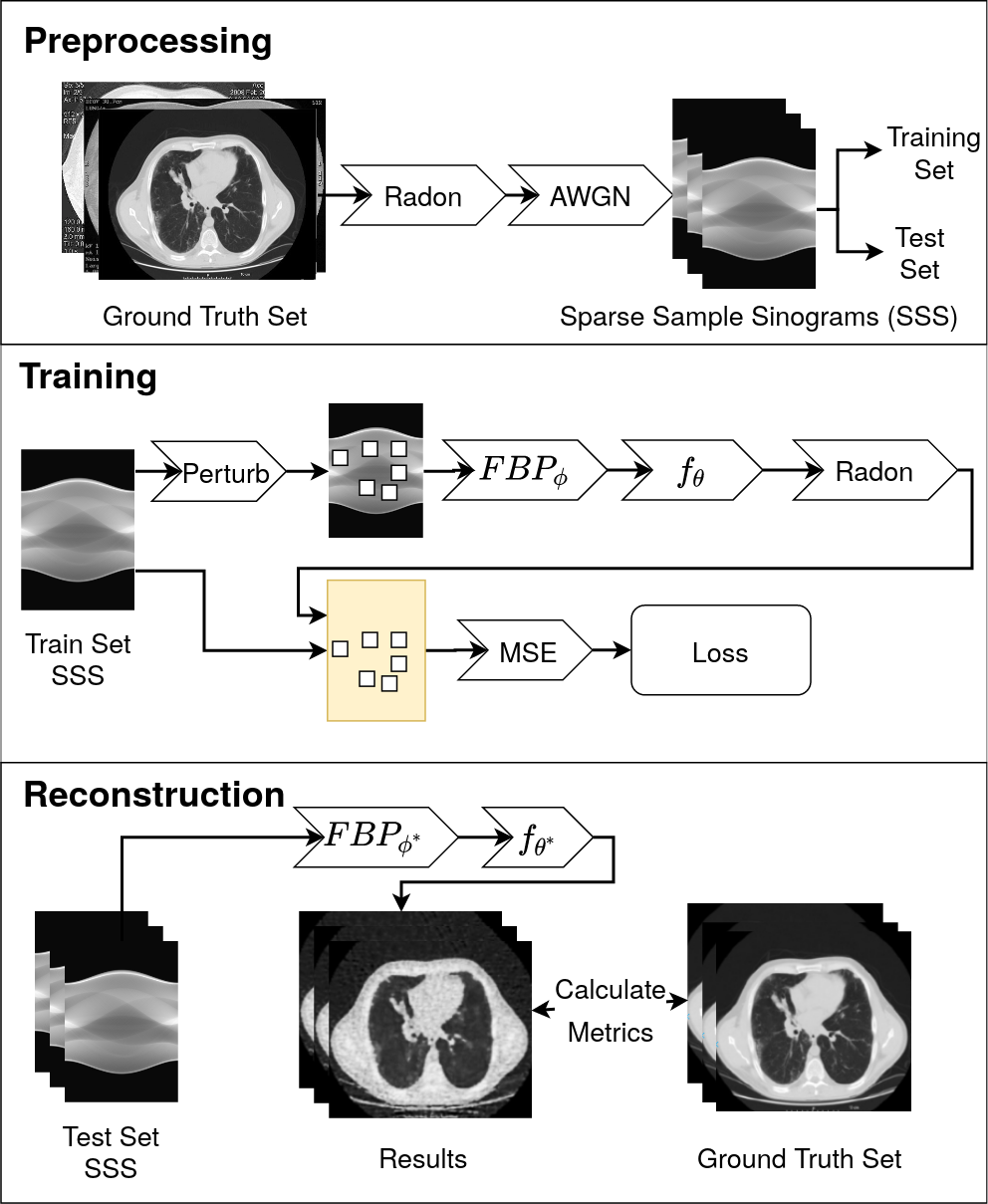}
        \caption{ Proposed working schema for self-supervised low-dose CT reconstruction.  }
        \label{fig:n2selfschema}
    \end{center}
\end{figure}

In this section, we explain how the self-supervision principle of Noise2Self method is extended for low-dose CT image reconstruction. Since FBP creates linearly dependent artifacts, it is not feasible to apply Noise2Self method for low-dose CT imaging in a straightforward sense. Regarding that, self-supervision should be exploited in such a domain as the requirements of Noise2Self method are met. In this manner, we designed a training scheme which learns a function to map low-dose CT images to standard-dose CT images without any standard-dose - low-dose image pairs dataset. To enable training only with low-dose CT data, we applied the J invariant principle of Noise2Self method in the projection domain where the noise can be modeled as element-wise independent. The suggested optimization process can be formulated as:

\begin{equation}
    \label{equ:n2s_ct}
    \{\theta^{*},\phi^{*}\}=\arg \min _{\theta,\phi} \mathbb{E}_{y}[\| A f_{\theta}(FBP_\phi(y_{Jc})))- y_{J} \|_{2}^{2}]
\end{equation}

\begin{figure*}[t]
    \includegraphics[width=1.0\textwidth]{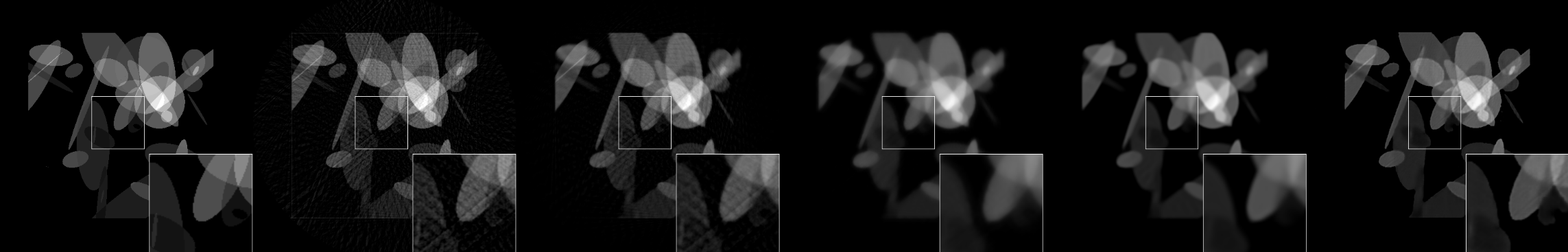}
    \caption{From left to right: ground truth, FBP, SART \cite{sart1984}, SART+TV \cite{sart_tv}, SART+BM3D \cite{bm3D}, the proposed method (learned self-supervised).}
    \label{fig:n2selfresult1}
\end{figure*}

\setlength{\tabcolsep}{2.5pt}
\begin{table*}[htbp]
    \centering
    \small
    \begin{tabular}{lllllllll}
        \hline
                                 & \multicolumn{4}{c}{Ellipses Phantoms} & \multicolumn{4}{c}{Real CT}                                                                                                                                                                                   \\
                                 & \multicolumn{2}{c}{32 view}           & \multicolumn{2}{c}{64 views}    & \multicolumn{2}{c}{32 view} & \multicolumn{2}{c}{64 views}                                                                                                                       \\
                                 & \multicolumn{1}{c}{PSNR}              & \multicolumn{1}{c}{SSIM}       & \multicolumn{1}{c}{PSNR}    & \multicolumn{1}{c}{SSIM}    & \multicolumn{1}{c}{PSNR}   & \multicolumn{1}{c}{SSIM}    & \multicolumn{1}{c}{PSNR}   & \multicolumn{1}{c}{SSIM}    \\
        \hline
        \small FBP               & $21.9\pm2.4$          & $0.53\pm0.08$          & $26.7\pm2.4$          & $0.70\pm0.08$          & $19.3\pm1.1$          & $0.43\pm0.05$          & $24.4\pm1.1$          & $0.62\pm0.06$ \\
        SART                     & $25.6\pm2.2$          & $0.77\pm0.06$          & $27.9\pm2.1$          & $0.84\pm0.05$          & $23.3\pm1.1$          & $0.75\pm0.05$          & $25.7\pm1.2$          & $0.86\pm0.03$ \\
        SART+TV                  & $26.2\pm2.1$          & $0.83\pm0.05$          & $28.1\pm2.0$          & $0.89\pm0.02$          & $24.6\pm0.8$          & $0.81\pm0.02$          & $26.1\pm0.8$          & $0.87\pm0.01$ \\
        SART+BM3D                & $26.4\pm2.1$          & $0.86\pm0.04$          & $\mathbf{28.8\pm2.0}$ & $0.92\pm0.02$          & $25.6\pm0.9$          & $0.87\pm0.02$          & $\mathbf{28.5\pm1.0}$ & $\mathbf{0.92\pm0.01}$ \\
        N2S Self Sup.            & $27.4\pm1.7$          & $0.90\pm0.02$          & $28.3\pm1.8$          & $0.92\pm0.01$          & $25.4\pm0.9$          & $0.89\pm0.01$          & $25.7\pm0.8$          & $0.91\pm0.01$ \\
        N2S Sng. Shot            & $25.7\pm1.8$          & $0.86\pm0.03$          & $26.6\pm1.7$          & $0.89\pm0.02$          & $23.9\pm0.8$          & $0.86\pm0.01$          & $24.6\pm0.8$          & $0.89\pm0.01$ \\
        N2S Learned              & $\mathbf{28.5\pm1.9}$ & $\mathbf{0.93\pm0.01}$ & $28.6\pm1.9$          & $\mathbf{0.94\pm0.01}$ & $\mathbf{25.9\pm0.9}$ & $\mathbf{0.92\pm0.01}$ & $25.6\pm1.0$          & $\mathbf{0.92\pm0.01}$ \\
        \hline
    \end{tabular}
    \caption{The average performance of the methods are given with PSNR and SSIM metrics respectively.}
    \label{table:benchmark}
\end{table*}

where $\mathbb{E}_{ y}$ is the expectation operator over $y$, $A$ is the forward operator of the inverse problem in our case Radon transform, $f_{\theta}$ is a deep neural network which is parameterized with $\theta$, $FBP_\phi$ is filtered back-projection reconstruction operator which has modified frequency-domain filter and is parameterized with $\phi$, $y_{Jc}$ is the perturbed form of the projections which is obtained by perturbing the $J^{th}$ subset of the projections ($y$), $y_{J}$ is the perturbed set of pixels. The working principle and training schema of the reconstruction is given in Fig. \ref{fig:n2selfschema}. The working principle can be examined in three parts:

\textbf{Preprocessing:} Ground truth images are converted to projections via Radon transform and contaminated with additive white Gaussian noise (AWGN). These sparsely sampled sinograms are split into two groups as training and test sets.

\textbf{Training:} The only input of training is sparsely sampled sinograms. The loss is calculated with self-supervision. First, projections are perturbed and used as the input of $FBP_{\phi}$ to calculate the initial image. The initial image is denoised with a deep neural network ($f_\theta$). The denoised image is transformed to the projection domain via Radon transform. The loss is calculated between back-projected measurements and the real measurements only at the pixels which are modified during the perturbation operation at the beginning of the reconstruction. In other words, only modified pixels are used to calculate the loss to satisfy the J invariant principle of Noise2Self method.

\textbf{Reconstruction:} During the reconstruction, the projections are used as the input of trained $FBP_{\phi^{*}}$ method without any perturbation to calculate initial reconstruction. After initial reconstruction, images are denoised with the trained deep neural network ($f_{\theta^*}$).

\section{Experiments}

The source code and the experiments are available at code repository\footnote{https://github.com/mozanunal/SparseCT}.

\subsection{Experiment Settings}

Deep lesion dataset \cite{yan2017deeplesion} and ellipses dataset were used as training data. SkipNet was selected as denoiser neural network architecture. The image resolution was selected as $512\mathbf{x}512$, and all projections were uniformly distributed between $0-2\pi$. The experiments are done with GPU (Nvidia RTX 2080 TI). Our method was compared with FBP, SART \cite{sart1984}, SART+TV \cite{sart_tv} and SART+BM3D \cite{bm3D}. The proposed idea was implemented in three different approaches:

\begin{figure}[t]
    \includegraphics[width=0.45\textwidth]{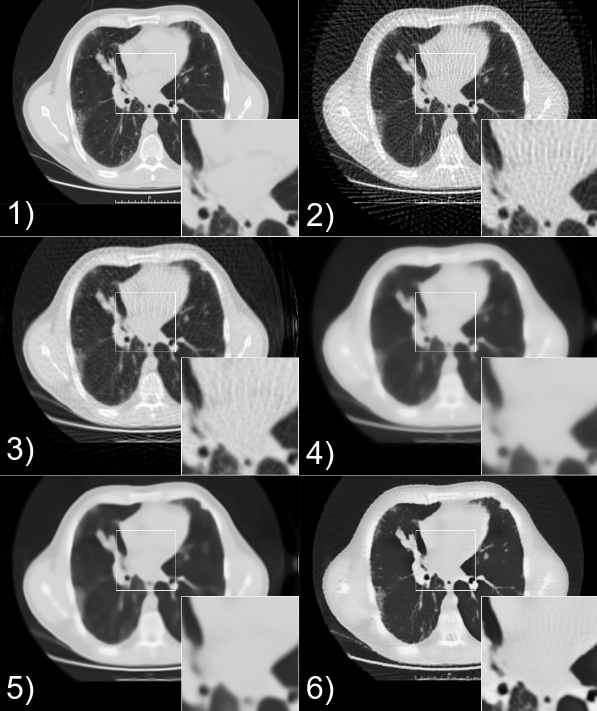}
    \caption{From top left to bottom right: ground truth, FBP, SART \cite{sart1984}, SART+TV \cite{sart_tv}, SART+BM3D \cite{bm3D}, the proposed method (learned self-supervised).}
    \label{fig:n2selfresult2}
\end{figure}

\textbf{Self-Supervised:} It was tested without any pre-learning process. A randomly initialized neural network was trained only with the single noisy image with the self-supervision principle which is given in the training part of Fig.\ref{fig:n2selfschema}. During the experiments, the complete reconstructions took $4000$ iterations with the learning rate $0.01$. The reconstruction time for this method is 240 seconds.

\textbf{Learned Single Shot:} The neural network was trained with a dataset and it calculates the denoised low-dose CT reconstruction in a single shot. In our case, the neural network was trained on $13000$ samples from ellipses dataset for $10000$ iterations batch size of $8$. The reconstruction time for this method is under 240 seconds.

\textbf{Learned Self-Supervised:} The neural network was trained with a dataset and it was fine-tuned with self-supervised training. This method generated the most successful results during the experiments. The reconstruction time for this method is 240 seconds.

To quantitatively analyze the methods, $10$ images from the ellipses dataset and $10$ images from the deep lesion dataset were selected to cover a more comprehensive part of different tissue intensity and feature scenarios. Images were reconstructed at different settings and the results are given in Table \ref{table:benchmark}.

\subsection{Results}

In Fig. \ref{fig:n2selfresult1}, the reconstruction of an image from the ellipses dataset can be examined. The reconstructed image by the proposed method created a sharper image with better noise redundancy performance.

Medical CT image results are given in Fig. \ref{fig:n2selfresult2}. The proposed method reconstructed the features more precisely while keeping the background smooth and clean. Although in some cases SART+BM3D method gives close or better quantitative results compared to the proposed method, it is clearly seen that the proposed method preserves the fine features more accurately.

\section{Conclusion}

In this study, we showed that self-supervised training can be a suitable candidate to solve the low-dose CT reconstruction problem. Its performance was evaluated with both real CT images and analytical phantoms. Since it does not require noise-free datasets, it is possible to use this method for numerous domains, particularly when collecting the noise-free data is challenging.

\apptocmd{\thebibliography}{\setlength{\itemsep}{0pt}}{}{}
\bibliographystyle{IEEEtran}
\bibliography{bibliography/ct,bibliography/ctdl,bibliography/dip,bibliography/dl,bibliography/metric,bibliography/n2n}

\end{document}